\documentstyle{article}
\begin{document}
\baselineskip 14pt
\thispagestyle{empty}
{\bf \Large Bond angle distribution in amorphous germania and silica}\\[1cm]

\noindent
{\large {\center J\"org Neuefeind}} \\ 
Hamburger Synchrotronstrahlungslabor HASYLAB at 
Deu\-tsches Elek\-tro\-nen\-syn\-chro\-tron DESY,
Notkestrasse 85, 22603 Hamburg, Germany 
{\large {\center K. -D. Liss}}\\
European Synchrotron Radiation Facility ESRF,
B. P. 220, 38043 Grenoble, France
\begin{abstract}
The distribution of Ge-O-Ge and Si-O-Si bond angles
$\alpha$ in amorphous germania and silica is 
re-determined on the basis of diffraction experiments. The bond angle $\alpha$
joining adjacent tetrahedra is the central parameter of any
continuous random network description (CRN) of these glasses.
New high energy photon diffraction experiments
on amorphous germania (at photon energies of 97 and 149 keV) are presented,
covering the momentum transfer 0.6-33.5\AA$^{-1}$. In
photon diffraction experiments on GeO$_2$ the contribution of the OO pairs
is very small. To obtain a similar information for amorphous SiO$_2$,
high energy photon diffraction experiments \cite{us} have 
been combined with neutron diffraction data
\cite{johnson,grimley} on amorphous silica
in order to eliminate the OO- partial structure
factor. With this technique it is
shown that the Si-O-Si angle distribution is
fairly narrow ($\sigma=7.5^{\rm o}$) and in fact 
comparable in width to the Ge-O-Ge angle distribution ($\sigma=8.3^{\rm o}$),
a result which differs from current opinion.
The narrower distribution found in this study are in
much better agreement to the determinations based on $^{29}$Si-MAS-NMR.
Among the various models relating the chemical shift to the bond angle,
best agreement is found with those models based on the
secant model.
Sharp components in the bond angle distribution can be excluded within the
reached real space resolution of 0.09\AA.
 
\end{abstract}

\vspace{1cm}

\section{Introduction}

Vitreous silica and germania are two prototype simple glasses. 
These two glasses consist of corner-sharing tetrahedra linked
to a continuous three-dimensional network. Numerous diffraction
experiments have been published, especially concerning amorphous
silica, of which the eldest date back to the very beginning of x-ray
diffraction \cite{vos}. But even before, some ideas about the structure
of glass already existed \cite{vvos}. The traditional theories of
glass structure are the crystallite theory and the continuous random network
(CRN) theory attributed to Lebedev \cite{lebe} and Zachariasen \cite{z}
respectively. Modern crystallite theory predicts fluctuations of the
 intermediate range order where some regions approach the atomic arrangement
of the crystal. Fluctuations of the intermediate range order are also 
allowed in the CRN model, so the difference between both models is mainly a
gradual one. As bond angles in crystals show only small variations
around discrete values, modern crystallite theory should lead to
sharp bond angle distributions superimposed on the broader distribution 
prevailing in the more disordered interconnecting regions. 
A more detailed discussion of these topics is given in the
review articles of Wright \cite{rewright} and Porai-Koshits \cite{pk}.

The central parameter of a CRN model of germania and silica
is the distribution of A-O-A bond angles (A=Si,Ge) joining adjacent
tetrahedra. The Si-O-Si angle distribution in amorphous
 SiO$_2$ is often believed to be much broader  than the corresponding
distribution in germania.
This can be traced back to 
the classical work of Mozzi and Warren \cite{m+w}.
Later some discussion came up about the
average Si-O-Si angle \cite{dasilva,coombs}, but not about the width
of the distribution. As the mean Ge-O-Ge angle of 133$^{\rm o}$ is close to
the one for planar 3-membered rings, Barrio {\it et al.} \cite{gal} recently
argued, that the 
narrower distribution in GeO$_2$ is an indication for the presence of 
such small, stiff ring-systems in a much larger extent than in SiO$_2$.
This hypothesis was inspired by the interpretation of IR-data of 
these glasses. 
Robertson {\it et al.} \cite{robert} computer relaxed the
Bell and Dean \cite{bd} CRN model, but were quite suspicious on their
result, as the Si-O-Si angle distribution  found was much narrower
than the Mozzi and Warren distribution.

The Si-O-Si angle can also be determined with $^{29}$Si-MAS-NMR spectroscopy.
The method is based on the assumption that a functional relation exists
between the chemical shift $\delta_{NMR}$ and the Si-O-Si bond angle.
This relation has
to be established experimentally in measuring the chemical shifts of 
a sufficiently large number of crystalline SiO$_2$ polymorphs with 
known Si-O-Si bond angles.
The first Si-O-Si angle determination by this technique was described by
Dupree and Pettifer \cite{nmr1}. This first distribution was claimed to be
very broad (FWHM 40$^{\rm o}$ !) and 
essentially flat between 140-155$^{\rm o}$. This result was later on revised,
being attributed to boron, sodium and aluminum contaminants present in the
investigated glass sample \cite{nmr2}. In this paper also the influence
of various assumptions on the relation between the chemical shift
and the bond angle is
discussed. In general, the bond angle distributions determined with  NMR
are found to be substantially narrower \cite{nmr2,nmr3} than the Mozzi and
Warren distribution, although in \cite{nmr3} a good agreement with Mozzi and 
Warrens result is claimed (This is certainly to be understood in comparison
to \cite{nmr1}).
MD simulation sometimes agree with the NMR results \cite{MD},
while other recent  simulations \cite{MDmw} find good 
agreement with Mozzi and Warrens result and a third group reported even a
broader distribution \cite{MDbr}.

Recently, diffraction experiments on amorphous systems 
(glasses: \cite{us}, molecular liquids: \cite{me}) with high energy 
photons as delivered from modern synchrotron radiation sources became feasible.
The term high energy photons will be used for photons in the energy
range of $\sim 80-200$ keV, i.e. of an energy about one order
of magnitude higher than that of x-rays traditionally used in 
angle dispersive measurements. As the photo-electrical absorption
decreases with energy approximately like $E^3$,
the absorption is reduced by three
orders of magnitude. Scattering therefore becomes the dominant process in
a large number of materials and the conditions of the experiment become
in general quite similar to that of 
a typical neutron diffraction experiment. The wavelength of the photons 
is quite 
short and consequently no principle limit exist on the maximum
momentum transfer attainable: The momentum transfer for back scattering   
with a photon energy of
100 keV is $Q_{max}=4\pi E/(hc)=
4\pi E \cdot 0.080658\mbox{keV}^{-1}{\rm \AA}^{-1}=
 101.4{\rm \AA}^{-1}$.
Considering only a more commonly used momentum transfer range the
scattering takes place through small or intermediate scattering angles.
This is advantageous for the use of small windows or flat area detectors, but
the first two characteristics - low absorption and high momentum transfer -
will be more important for the present study. 
For a description of the use of high energy photons in general - i. e. not
restricted to amorphous substances - see \cite{schnei95}.

This article will focus on the Si-O-Si and the Ge-O-Ge bond angle
distributions and its comparison. The 
data on amorphous germania are new and the experiment is
described in the following section. The experiment on vitreous silica 
at the DORIS-III synchrotron 
has been described previously \cite{us}. 
The neutron diffraction data on amorphous
silica have been taken from \cite{johnson} and \cite{grimley}.

\section{Experiment on germania}
The vitreous germania has been prepared from 99.999\% pure GeO$_2$ 
(Strem-Chemicals, Newburyport, MA, USA) by melting
the crystalline material. The melt was kept for 8 h at 1473 K at atmospheric 
pressure (melting point 1383 K). The furnace was then switched
off and the sample cooled to 900 K in about 10 min. 
The resulting sample was clouded as a result of the
inclusion of small gas bubbles but no traces of crystallinity were left.

The diffraction experiment on vitreous germania has been carried out at
the high energy 
beam-line ID15A of the European Synchrotron Radiation Facility ESRF.
The Synchrotron was 
operated in 16 bunch mode with an average current of about 70 mA
and beam life-times of 
about 16 h. The beam-line is designed to operate alternatively with a
permanent magnet multipole wiggler or with a super conducting 
wavelength shifter \cite{esrf}. The former only is commissioned 
at the moment and thus has been used for the experiment. Its gap was 
maximal closed to 
20.3 mm leading to a critical energy of 43 keV. To remove the 
heat load, lower energies are 
cut by a cooled 4 mm aluminum filter. The beam size can be chosen by a set 
of water-cooled 
secondary slits in front of the monochromator position in the optics 
hutch II \cite{esrf}. 

A perfect (220) Si crystal in Laue geometry was used for monochromatization
and mounted on 
a bending device. To reduce background in the experimental hutch, 
the white beam was stopped in the optics hutch II. 

A dedicated two axes diffractometer at the high energy 
beam-line does not exist. 
Instead the sample was mounted on the analyzer stage of 
the triple axes diffractometer. Here, 
the scattering angle is chosen by a translation of the detector 
perpendicular to the white beam 
axis. This geometry allows to select a maximum scattering angle of 2$\theta$
= 26$^{\rm o}$. The solid state detector was equipped with a 30 mm thick
germanium crystal and has a 
typical energy resolution of about 1 keV. Discriminating the energy around 
the photon energy of the monochromatic beam reduces the background scattered 
from the white beam. The intensity in 
the first sharp diffraction peak of the sample was around $5\cdot 10^4$ counts 
per second. Further increase 
of the intensity would raise considerable dead-time problems.
In order to cover a large momentum transfer range two different photon 
energies of 97.58 keV and 149.80 keV were chosen. 
The monochromator was bent for the higher energy 
setup in order to compensate for the intensity drop.

 The energy of the primary beam was determined by observing
the Bragg peak position of a second perfect (220) Si crystal mounted 
on the monochromator position of the triple-axes diffractometer for both,
dispersive and non-dispersive geometry. 

Due to the excellent beam stability of the ESRF synchrotron,
normalizing the data on the beam current was found to be more
convenient than using an NaI monitor, which was installed for control
purposes. The latter observes the air scattering produced 
by the incoming beam and was too sensitive to the hutch background
as its energy resolution is too poor. 
The problems associated with beam fluctuations described in \cite{us}
are mostly reduced in the present study.
 In order to minimize generally the effect
of drifts the  data acquisition was organized as a sequence 
of several short scans.

The data were subsequently corrected for background, absorption,
multiple scattering, polarization of the incoming beam, 
the variation of the solid angle seen by the detector as 
a result of the changing distance between the sample and the detector
(tangential movement, see above) and
the increased detection efficiency for inelastically scattered
photons.
$\mu r$ is 0.25 at 149 keV (with $\mu$ being the total absorption
 coefficient and $r$ the radius of the sample). Air scattering is 
the dominant background contribution, especially at low Q's. 
The correction procedure was described in detail in \cite{us}.
The resulting fully corrected and normalized data are presented in Fig. 1. 

\section{ Data analysis}

The usual definition of the structure functions in the neutron
and photon case are:
\begin{eqnarray}
S(Q)=\frac{ \left( \displaystyle {\frac{d\sigma}{d \Omega}} \right )_{dist}^n}
{(\sum_{uc} b_i)^2} =\sum_{jk} \frac{b_j b_k}{(\sum b_i)^2} s_{ij}  &&
i(Q)=\frac{ \left( \displaystyle {\frac{d\sigma}{d \Omega}} \right )_{dist}^x}
{(\sum_{uc} f_i)^2} =\sum_{jk} \frac{f_j f_k}{(\sum f_i)^2} s_{ij}
 \label{sofq} 
\end{eqnarray} 
where $ \left (\frac{d\sigma}{d \Omega} \right )_{dist}$ is the distinct
(interference) differential cross section, $S$ and $b_i$ are the structure
function and the scattering lengths in the neutron case, $i$ and $f_i$ are
the structure function and the form-factors in the photon case, the sum
is extending over the unit composition. $S(Q)$ and $i(Q)$ are weighted sums
of the partial structure factors $s_{jk}$. The $s_{jk}$ are, of course,
 the same in neutron and photon diffraction. The weighting 
coefficients in the photon case are constant in $Q$ only if one
assumes that the form factors of the atoms present differ by
a scaling factor only - this is 
the K-approximation introduced by Warren, Krutter and 
Morningstar \cite{gmorning}. 
A combination of neutron and photon data given by:
\begin{equation}
d_{j'k'}(Q)= f_{j'} f_{k'}
 \left( \displaystyle {\frac{d\sigma}{d \Omega}} \right )_{dist}^n - b_{j'}
 b_{k'} \left( \displaystyle {\frac{d\sigma}{d \Omega}} \right )_{dist}^x
= \sum_{jk \neq j'k'} (f_{j'} f_{k'} b_j b_k - b_{j'} b_{k'} f_j f_k )
s_{jk}
\label{eq-fod}
\end{equation}
eliminates one particular
partial structure factor $s_{j'k'}$ from the diffraction
pattern without any approximation.
This procedure is analogous to the isotope substitution
technique in neutron diffraction,
but has the advantage that the same sample can be used.
Prerequisites for a successful separation are a good knowledge of the
form-factors and of the neutron scattering lengths.
The neutron scattering lengths used in the
 analysis are taken from \cite{koester},
the form-factors from \cite{hubbell1,hubbell2}.

The three differences which are possible for SiO$_2$ 
according to Eq. (\ref{eq-fod}) are shown in Fig.
2. The contribution of the first two peaks in real space
- the SiO-  and OO- first shell - calculated
according to \cite{grimley} is  
indicated. It is apparent from the figure that the scattering pattern is 
dominated by these peaks above 15 \AA$^{-1}$. It is also apparent that
a systematic error is present between 16 \AA$^{-1}$ and 20 \AA$^{-1}$, 
probably caused by the splicing of two scans in that region \cite{us}.
This can also be seen in Fig. 3 where the contribution of the
first two peaks has been subtracted.

A useful real space correlation  function which can be used
for the analysis of amorphous materials is (in the neutron case):
\begin{equation}
T(r) = \frac{1}{2 \pi^2 \rho} \int_{Q_{min}}^{Q_{max}} Q S(Q) M(Q)
 \sin(Qr)dQ + r
=\sum_{jk}  \frac{b_j b_k}{(\sum b)^2} t'_{jk}(r), 
\end{equation} 
where $M(Q)$  is an optional modification function.
$T(r)$ is related to $g(r)$ - the correlation function which is
more commonly used in the discussion of the structure of 
liquids - simply by $T(r)=r g(r)$.
The experimental component correlation functions $t'$ are related to the real
 component correlation functions $t$ via a convolution with a peak shape
function \cite{m+w,scho}. The peak shape function is determined by the
modification function used and - in photon diffraction - 
the non-constant weighting factors of the partial structure factors.
The modification function used here is the rectangular function 
(all data points are equally weighted for the Fourier transform):
\begin{equation}
M(Q)=\left\{
 \begin{array}{r@{\quad:\quad}l} 0& Q<Q_{min}\\1&Q_{min}<Q<Q_{max}\\
0&Q>Q_{max} \end{array} \right.
\end{equation}
This is to the authors opinion a better solution than weighting  
parts of the measured data with a modification function as e. g. the one
proposed by Lorch \cite{lorch} as long as the implications on
$T(R)$ are noticed.

For the differences given by Eq. (\ref{eq-fod}) the Fourier transform has been
calculated according to:
\begin{equation}
\label{eq-norm}
T_{diff}(r) = \frac{1}{2 \pi^2 \rho (f_{j'} f_{k'} b_{j''} b_{k''} -
 b_{j'} b_{k'} f_{j''} f_{k''} )N} \int_{Q_{min}}^{Q_{max}}
Q d_{j'k'}(Q) M(Q) \sin(Qr)dQ + r
\end{equation}
The division by 
$ f_{j'} f_{k'} b_{j''} b_{k''} - b_{j'} b_{k'} f_{j''} f_{k''}$
removes the dependence of the peak shape function from the non-constant
weighting factors for one of the partial correlation functions (for $j''k''$).
$N$ is defined by:
\begin{equation}
N=\sum_{jk \neq j'k'} 
\frac{f_{j'}(0) f_{k'}(0) b_j b_k - b_{j'} b_{k'} f_j(0) f_k(0)}
{f_{j'}(0) f_{k'}(0) b_{j''} b_{k''} - b_{j'} b_{k'} f_{j''}(0) f_{k''}(0)}
\end{equation}

Whenever comparisons between a model and the experiments are presented,
a double Fourier-transform has been carried out: The model partial correlation
functions $t(r)$ are transformed into $Q$-space:
\begin{equation}
Qs(Q)= 4\pi \rho \int_0^\infty t(r) \sin(Qr) dr
\end{equation}
multiplied by the appropriate weighting factors, summed and back transformed
to $r$-space with the same Q-interval than the experimental data. In short,
the model correlation functions are convolved 
with the appropriate peak shape function before
comparison. The agreement with the experiment is judged according to 
\cite{grimley} by calculating:
\begin{equation}
\chi^2=\frac{1}{\int_{r_{min}}^{r_{max}}T^2_{exp}(r) dr}
\int_{r_{min}}^{r_{max}} (T_{exp}(r)-T_{model})^2 dr
\label{eq-chi}
\end{equation}

Models have been developed using a MC algorithm. The A-O bond length (with
A=Si,Ge) is assumed to have a Gaussian form:
\begin{equation}
V(r_{A-O})=K \cdot \exp \left [-\frac{(r_{A-O}-r_0)^2}{2 \sigma_{r}^2}\right ],
\label{eq-dist}
\end{equation}
with K being a scaling factor, $r_{A-O}$ the bond length,
$r_o$ and $\sigma_r$ having its
usual significance. The bond angles A-O-A and O-A-O have been 
modeled by distributions of the form:
\begin{equation}
 V(\alpha)=K\cdot \exp \left [ -\frac{(\alpha-\alpha_0)^2}{2 \sigma^2} \right ]
\cdot \sin(\alpha),
\label{eq-ang}
\end{equation}
with $K$ again a scaling factor, $\alpha$ the A-O-A bond angle,
$\alpha_0$ and $\sigma$ are the free parameters of the distribution.
The O-A-O bond angle will be called $\beta$ in the following, and 
$\alpha$ has to be replaced by $\beta$ then in the above equation.
The dihedral angle distribution has been given the form:
\begin{equation}
 V(\delta)=K\cdot \sum_{i=1}^5 \exp 
\left [ -\frac{[\delta-\delta_0\cdot(i-2)]^2}
{2 \sigma_\delta^2} \right ],
\label{eq-dihed}
\end{equation} 
where $\delta$ is the dihedral angle A-O-A-O,
 $\delta_0$ has been set to 60$^{\rm o}$, K is a scaling factor and
$\sigma_\delta$ is a variable parameter. This leads to distributions 
having the necessary threefold symmetry favoring eclipsed conformation
(In the limit $\sigma_\delta \rightarrow \infty$ an uniform dihedral 
angle distribution is reached). Assuming independence of the
above distributions the short to intermediate range order is
specified: The A-O, O-O, A-A first and second shell contributions. These
distributions are calculated and compared with the experiment
via the above described formalism.

\section{Simulation}
To calculate the distance distributions  from the distributions (\ref{eq-dist})
-({\ref{eq-dihed}) chains O-A-O-A-O-A of A and O atoms were generated.
Beginning the chain from one side - say with the O- atom 
a (normally distributed)
random number determines the position of the next A- atom.
The distribution (\ref{eq-dist}) is, hence, fulfilled in this step. 
The position of the next O- atom is determined by
two random numbers (representing distribution (\ref{eq-dist})
and (\ref{eq-ang})),
the following A- atom by three random numbers and so on.
As only four- body correlations and no cross correlations are considered 
a maximum of three random numbers is needed for positioning an atom in 
the chain. Each chain yields five times a A-O first shell distance,
 two times a O-O and A-A first shell, three times a
A-O second shell and one times a O-O and A-A second shell distance.
$10^5$ chains were generated for each model.

This MC approach has the advantage over the analytical
approach followed in \cite{us} to allow a much more flexible
choice of distributions, while still being computationally
very cheap (few seconds of CPU time). It has the disadvantage
that the analytical form of the distributions has to be
imposed in fore-hand. It is easy to overcome this problem, following the
ideas of reverse Monte- Carlo (RMC) simulations \cite{RMC}. 
First one generates chains with quite wide distributions of next neighbor
distances, bond angles and dihedral angles. The generated chains 
are accepted or
rejected using the Metropolis criterion, if these chains improve
or deteriorate the agreement of the total model (the ensemble
of chains accepted so far) with the experiment.
This technique is by orders of magnitude more time consuming as
it needs the double Fourier transforms described in the preceding chapter
to be executed for each chain generated. The next more
complex simulation would be a full RMC simulation of the system:
This introduces an additional information into the simulation process
, i. e. the macroscopic density, but on the other hand - a disadvantage 
in a system which is in reality of course aperiodic -
periodic boundary conditions. A RMC simulation for SiO$_2$ has already
been carried out
(including information from both neutron and photon diffraction
experiments)
\cite{Natur} but it leads to structures which are apparently unphysical
on various counts. One reason is perhaps, that  a "good" structure can not
be reached within the computer time available.

\section{Results and Discussion} 
$T_x(r)$ for amorphous germania are shown
in figure 4a.
Looking closer at the GeO next neighbor peak
in Fig. 
4b, one recognizes that the peak width decreases and
the peak height increases 
steadily with increasing $Q_{max}$: The peak form is largely determined
by the truncation effect and the Fig.
4b gives, hence, an idea 
of the real space resolution of various techniques: $Q_{max}=14 {\rm \AA}^{-1}$
corresponds to a conventional x-ray diffraction experiment,
$Q_{max}=23 {\rm \AA}^{-1}$ to the D4b neutron diffractometer of the ILL 
(Grenoble) and finally $Q_{max}=33 {\rm \AA}^{-1}$ can be easily reached with
high energy photons. This behavior also proves, that the region 
$23\mbox{\AA}^{-1}<Q<33\mbox{\AA}^{-1}$ still contains valuable information
about this peak.
The GeGe next neighbor distance in Fig. 4c, in contrary,
does not change its form for
 $Q_{max}> 23 $\AA$^{-1}$.
The GeGe peak has consequently reached its
natural form and extending the integration to higher Q values  can only
obscure the peak shape by inclusion of the high Q noise. From the position and
 the width
of the GeO and the GeGe next neighbor peaks one can deduce the bond
angle distribution.
$\alpha_0=133.0
^{\rm o}$
and $\sigma=8.3
^{\rm o}$. This is in good agreement
 with an earlier x-ray structure determination
\cite{lbw} using Mo-K$^\alpha$ x-rays (17.4 keV),
ending with the more qualitative statement: ''[...] {\it 
the average Ge-O-Ge angle is {\rm 133}$^{\rm o}$. The distribution
of Ge-O-Ge angles is fairly narrow but cannot be determined
quantitatively with data of the present resolution.}'' 
$\chi$ defined by Eq. (\ref{eq-chi}) is 2.0\% in the range 
$2.5\rm{\AA}<r<4.0\rm{\AA}$ and 2.8\% in the range from $0<r<4$\AA.
$\chi^2$ does not change significantly, when the $\sigma_\delta$ parameter
is changed from infinity (randomly distributed dihedral angles) to 38$^{\rm o}$.
Decreasing $\sigma_\delta$ further deteriorates the fit. In the limit
of pure staggered or pure eclipsed conformation ($\sigma_\delta \rightarrow 0)
$ a $\chi $ of 6.5\% and 5.9\% is reached. The final model parameters
are collected in Tab. \ref{tab-mp}. The model is compared to the experimental
$Qi(Q)$ in Fig. 5. 

There is no indication for a sharp component in the bond angle distribution,
which should be expected, if a significant volume fraction of the 
glass would reach crystalline order.

The question arises, whether the connection of the tetrahedral units is
really that different in germania and silica. To answer this question
a reanalysis of previous high energy photon experiments \cite{us} was
started, combining them with neutron diffraction data \cite{johnson,grimley} 
in the way 
described above. The most interesting of the first order differences
$d_{j'k'}$ presented above is the one where the OO contribution is removed.
The Fourier-Bessel
transform is shown in Fig. 
6. $Q_{max}$ has been chosen
to be 16\AA$^{-1}$ for the reasons outlined in the previous section.
The data are normalized according to Eq. \ref{eq-norm} such that
the SiSi partial structure factor has a constant weighting factor.
Assuming again 
Si-O-Si bond angle distributions as in Eq. \ref{eq-ang}, broad distributions
with $\sigma>10^{\rm o}$ are clearly unacceptable. From that it follows
that the assumption of isotropic dihedral angles is not justified. The 
geometric relations are such, that isotropic dihedral angles \underline{and}
narrow bond angle distributions lead to a shoulder or even a distinct peak
on the low r side of the Si...O second neighbor peak. Some evidence
was already collected by earlier authors on the non-randomness of the dihedral
angles, but the literature is controversial in this point.
So, e. g. Galeener \cite{galle1}
(from sterical arguments) and Evans {\it et al.}  (from a comparison of various
models with neutron diffraction data) suspected the dihedral angles to
be non-randomly distributed. Gaskell {\it et al} \cite{gt} found 
a non uniform dihedral angle distribution in his relaxed version of the
Bell and Dean \cite{bd} model, while Gladden \cite{gladden} 
on the other hand found uniform dihedral angles in her models.
The best agreement
with the experiment is reached with $\alpha_0= 148.3
^{\rm o}$, $\sigma=7.5
^{\rm o}$ and $\sigma_\delta$=27$^{\rm o}$.
 The final parameters of the model are collected in Tab. 1

The interpretation of the
NMR line-shape of silica glass is based on  the knowledge of a relation
between the chemical shift and the Si-O-Si angle. Four such relations
are discussed in reference \cite{nmr2}: The secant \cite{ang1}, the linear 
\cite{ang2} ,the
point charge \cite{ang3} and the s-hybridization \cite{ang4} model.
 The secant and the
s-hybridization model produce almost identical Si-O-Si angle distributions
and are not considered separately. The resulting bond angle
distributions are compared in Fig. 7 with each other, and with
the distribution obtained in this work.
$\chi^2$ as a function of $r_{max}$ is shown in Fig. 6.
 While the linear
model gives a bad agreement with the combined neutron and photon
data, the secant
(with the data of Pettifer {\it et al.}) and point charge model 
give a comparable fit quality , but the lowest $\chi$ of all 
NMR distributions has the bond angle distribution given by Gladden {\it et al.}
These authors also used the secant model to derive their bond
angle distribution. 
It is interesting to note, that the tail to large angles present in
many published Si-O-Si distributions is not necessary to explain
the diffraction data. The $\chi$ values reached by various distributions 
are reported in Tab. 2.
All experimenters, of course, used different samples - with 
different concentrations of impurities and defects, different quenching
conditions {\it etc.}
- and this may also play a significant role at this level. 
It would be interesting to go an alternative route: The determination of 
the functional relation between chemical shift and bond angle by  investigating
the same glass
sample with neutron and photon scattering and $^{29}$Si- MAS- NMR.

Several authors devoted 
special attention to the first sharp diffraction peak (FSDP)
(see the discussion given by Elliot 
\cite{Elliot}). It is noted, that the FSDP vanishes almost completely
when the SiO partial structure function is eliminated (Fig. 2). Probably the
OO- and the SiSi- partial cancel as its weighting factors have different
sign. Furthermore it is noted that the FSDP can not even be recognized any more
if the contribution of the short range order is subtracted (Fig. 3). It is thus 
concluded that the FSDP by itself has very little structural significance
as it is more or less arbitrarily
composed of the weighted partial distribution functions and contributions of
short and medium range order. This will be
even worse in glasses of more complicated composition.

The experimental $QS(Q)$ and $Qi(Q)$ respectively are compared
to the model in Fig.
8. The specified elements of
the short and intermediate range order can describe both the neutron and photon
diffraction pattern down to about 2.5 \AA$^{-1}$. The contribution
of the different first and second neighbor distances to 
$T_x(r)$ is shown in Fig. 9
and can be compared to
 Mozzi and Warrens Fig. 4 \cite{m+w}.

Finally, one should discuss what possibilities exist for
a further disentangling of the partial structure factors.
There is very little that  can be done for silica. There are 
no suitable isotopes for oxygen, the scattering length difference
between $^{28}$Si and $^{29}$Si is small (0.6fm) and $^{29}$Si is
quite expensive. Anomalous diffraction and EXAFS spectroscopy are
hindered by the low energy of the Si-K edge (1.8 keV). This is,
why a combination of photon and neutron data is especially
valuable in this case. In contrast, there are suitable
Ge-isotopes for an isotope substitution experiment 
($^{70}$Ge and $^{73}$Ge) \cite{koester}.
So, GeO$_2$ could be a promising candidate to get a very detailed
structural picture of a simple tetrahedral glass. 

\section {Conclusion} 

Combining neutron and photon diffraction 
can yield additional information
on such well studied  systems as the simple glasses. The ideal counterpart
to modern neutron instrumentation are high energy photons in the energy 
range from $\sim$ 80-200 keV, which can be delivered from modern 
synchrotron radiation sources. The main advantages which come into play in the
present study is the large accessible Q-range and the low absorption.

Due to the large accessible Q-range the GeGe next neighbor peak can
be observed without any  truncation effect. This leads to an accurate
Ge-O-Ge bond angle distribution, the central parameter of
a CRN description of the glass. There is no indication of a 
sharp component in the bond angle distribution, which could be expected,
if a significant volume fraction of the glass would reach crystalline
order.

Combining neutron and photon results can lead to an isolation 
of the SiSi first neighbor peak in SiO$_2$, too. From this, it can
be deduced  that the very broad Si-O-Si distribution calculated 
by Mozzi and Warren is not acceptable. 
This broad distribution is a consequence of the assumption of
randomly distributed dihedral angles. Likewise the photon data used here would
lead to a broad Si-O-Si bond angle distribution - when interpreted
alone without combination with neutron data and using this wrong
assumption \cite{us}.
Only in combination with neutron data, it becomes clear that preference
should be given to narrower distributions and the width of
the Si-O-Si distribution is indeed within the experimental accuracy found to be
identical to the width of the Ge-O-Ge angle distribution.
From the bond angle distributions determined
by $^{29}$Si-MAS-NMR the one given by Gladden {\it et al.}\/ gave the best
fit to the combined neutron and photon data.
The Si-O-Si distributions determined form diffraction and from NMR
measurements are in much better agreement, when 
neutron and photon diffraction are combined. 
Among the models relating the chemical shift to the bond angle
the secant model agrees best with the diffraction data.
These narrower distributions
demand in turn a non-uniform distribution of
the dihedral angles, that is to say a preference of the 
staggered conformation. The bond length, the Si-O-Si angle and
the dihedral angle are likely to be correlated among each other and in 
neighboring units, but no evidence can be taken from
diffraction experiments alone on these correlations.

\section{Acknowledgments}
Financial support of the Deutsche Forschungsgemeinschaft DFG under
Grant No. Ne584/1-1 is gratefully appreciated.

\clearpage 
\section{Figures}
{\center Figure 1: The fully corrected and normalized data 
(high energy photons,GeO$_2$)}

Full line:  experimental intensity; dashed line: isotropic part
 of the scattering intensity (Self scattering + Compton scattering)

The insert is a zoom at the high-Q region.

{\center Figure 2: The first order differences $d_{j'k'}$ in amorphous silica}

From top to bottom:  $Qd_{SiSi}(Q)$, $Qd_{OO}(Q)$, $Qd_{SiO}(Q)$.
Solid line: experimental 
$Qd_{j'k'}(Q)$, dashed line contribution of the first two peaks (SiO-,OO-)
in real space.

{\center Figure 3: The first order differences in amorphous silica,
contributions of the first two peaks in real space removed}

From top to bottom:  $Qd_{SiSi}(Q)$, $Qd_{OO}(Q)$, $Qd_{SiO}(Q)$. 

{\center Figure 4: $T_x(r)$ for vitreous germania}

a) $T_x(r)$ for vitreous germania, $Q_{max}=23 {\rm \AA}^{-1}$

b) and c) Closer look at the GeO- and the GeGe-peak respectively:
solid line $Q_{max}=33 {\rm \AA}^{-1}$, dashed line $Q_{max}= 23{\rm \AA}^{-1}$,
dots $Q_{max} = 14{\rm \AA}^{-1}$

{\center Figure 5: 
Comparison of the model distribution for GeO$_2$ with experiment}

Solid line: experimental  $Qi(Q)$ , dashed line: model $Qi(Q)$

{\center Figure 6: $T_{OO}(r)$ and $\chi^2$ from the various distributions}

Upper window:$T_{OO}(r)$; lower window $\chi^2$ ($\chi^2$ is not normalized
to $\int T_{exp}^2(r)$ here).

Full line: experiment,
 dashed triple dotted:
Mozzi and Warrens distribution \cite{m+w},
1. dotted line: Best fit of Eq. (10),
2. dashed dotted line: NMR model of ref. \cite{nmr3},
3-5: NMR models discussed in ref. \cite{nmr2};
3. linear model (not shown)
4. dashed double dotted line:
NMR data of ref. \cite{nmr2}, point charge model;
5. dashed line 
NMR data of ref. \cite{nmr2}, secant model;
6. triple dotted line:
best fit of Eq. (10) to the NMR distribution of ref. \cite{nmr3}

{\center Figure 7: Various models for the Si-O-Si bond angle}

1. dotted line: Best fit of Eq. (10) to the combined photon and neutron data, 
2. dashed dotted line: 
NMR result from ref. \cite{nmr3}, based on the secant model; 
3-5: NMR models discussed in ref. \cite{nmr2};
 3. dashed triple dotted line: linear model,
4. dashed double dotted line: point charge model,
5. dashed line: secant model. 
6. triple dotted line: best fit of Eq. (10) to the NMR
result from ref. \cite{nmr3} 

{\center Figure 8: Comparison of the model for SiO$_2$ with the experiment}

Solid line: experimental $QS(Q)$ and $Qi(Q)$ respectively;
dashed line: model $QS(Q)$ and $Qi(Q)$ respectively.
a) photon diffraction \cite{us}, b) neutron diffraction \cite{johnson,grimley}

The neutron data were available as spline fit to the experimental data only.

{\center Figure 9: 
First and second shell contributions to $T_x(r)$}

Solid line: experimental $T_x(r)$ and Fourier transform of the residue
with $Q_{max}=10.6{\rm \AA}^{-1}$, dashed line: SiO- and OO- first
neighbor, dotted line: SiSi- first shell,
dashed dotted: SiO- second shell, triple dotted line: OO second shell,
double dotted line SiSi- second shell contribution

\begin{table}[h]
\caption{Structural parameters for SiO$_2$ and GeO$_2$}
{\center \begin{tabular}{|c|ccccccc|}
\hline
&$r_{AX}$& $\sigma_r$&$\alpha_0$&$\sigma_\alpha$&$\beta_0
$&$\sigma_\beta$&$\sigma_\delta$\\ \hline
GeO$_2$&1.73&.0415&133.0&8.3&109.47$^1$&4.2$^2$&38\\
SiO$_2$&1.605$^3$&.0493$^3$&148.3&7.5&109.47$^1$&4.2$^3$&27\\\hline
\end{tabular}} 

\noindent
Distances in \AA, angles in degree.

\noindent
1: Not refined, $\beta_0$ is taken from the ideal tetrahedron 

\noindent
2: Not refined, taken to be equal to SiO$_2$

\noindent
3: Not refined, taken from reference \cite{grimley}
\label{tab-mp}
\end{table}
\begin{table}
\label{tab-chi}
\caption{$\chi$ reached from various bond angle models}
{\center \begin{tabular}{|c|c|c|}
\hline
\multicolumn{2}{|c|}{This work}&0.84\%\\ \hline
\multicolumn{2}{|c|}{Gladden {\it et al.}\cite{nmr3}}&1.4\%\\ \hline
&secant model&2.26\%\\
Pettifer {\it et al.}\cite{nmr2}&point charge model&2.20\%\\
& linear model& 3.41\%\\ \hline
\end{tabular}}
\end{table}
\end {document}
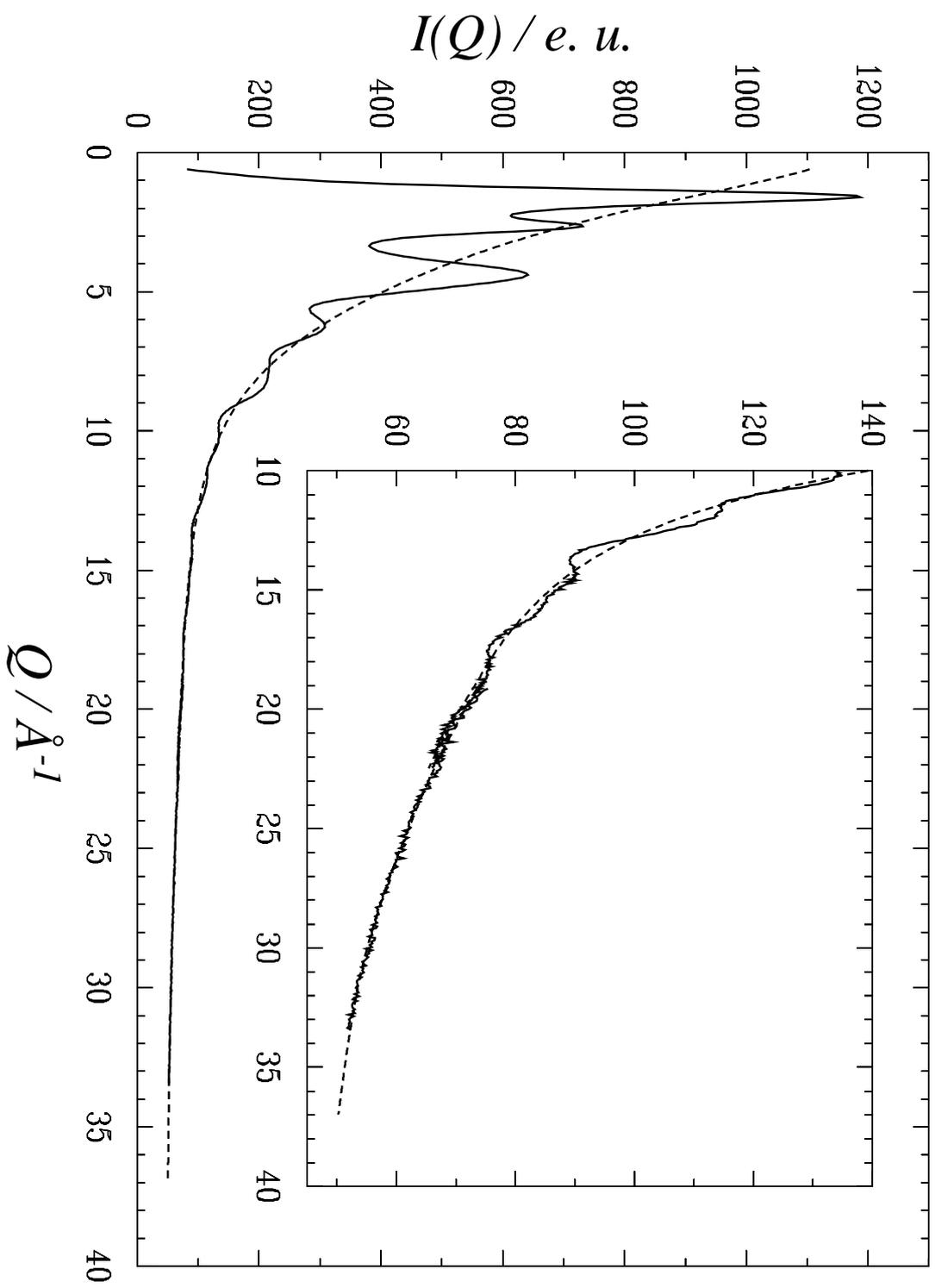
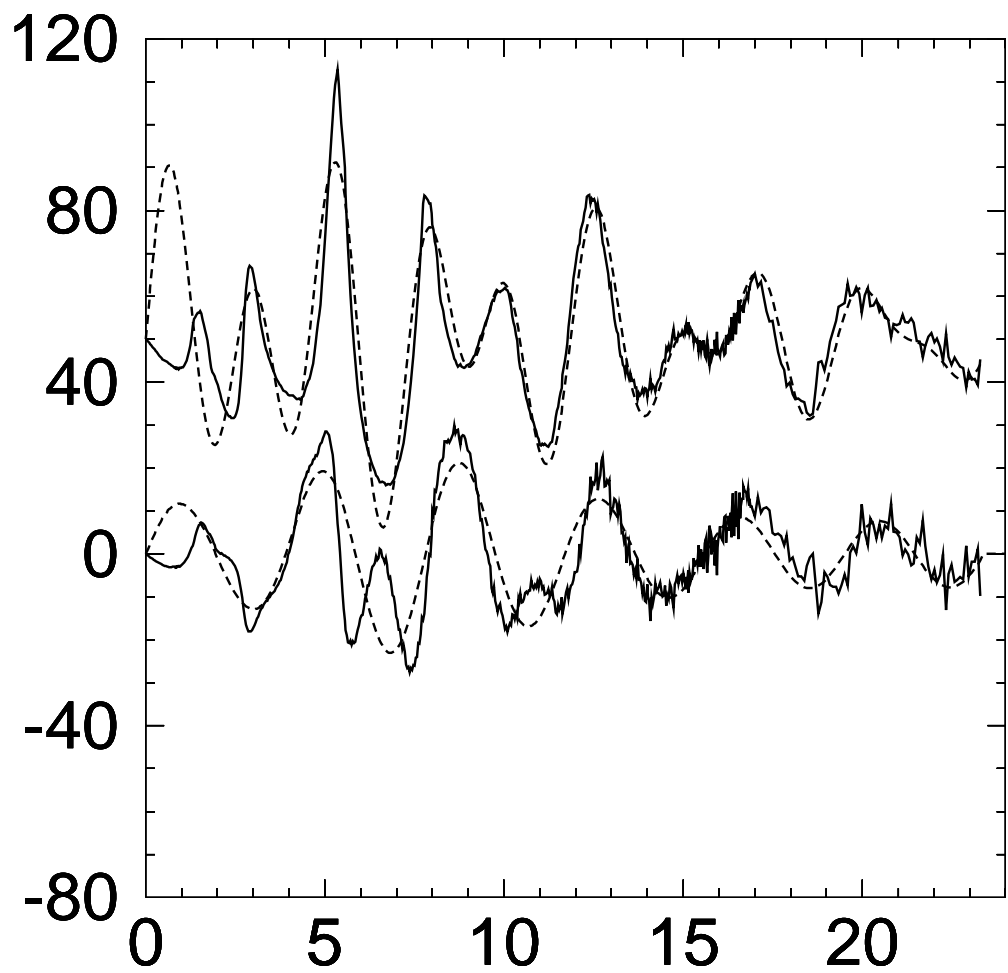
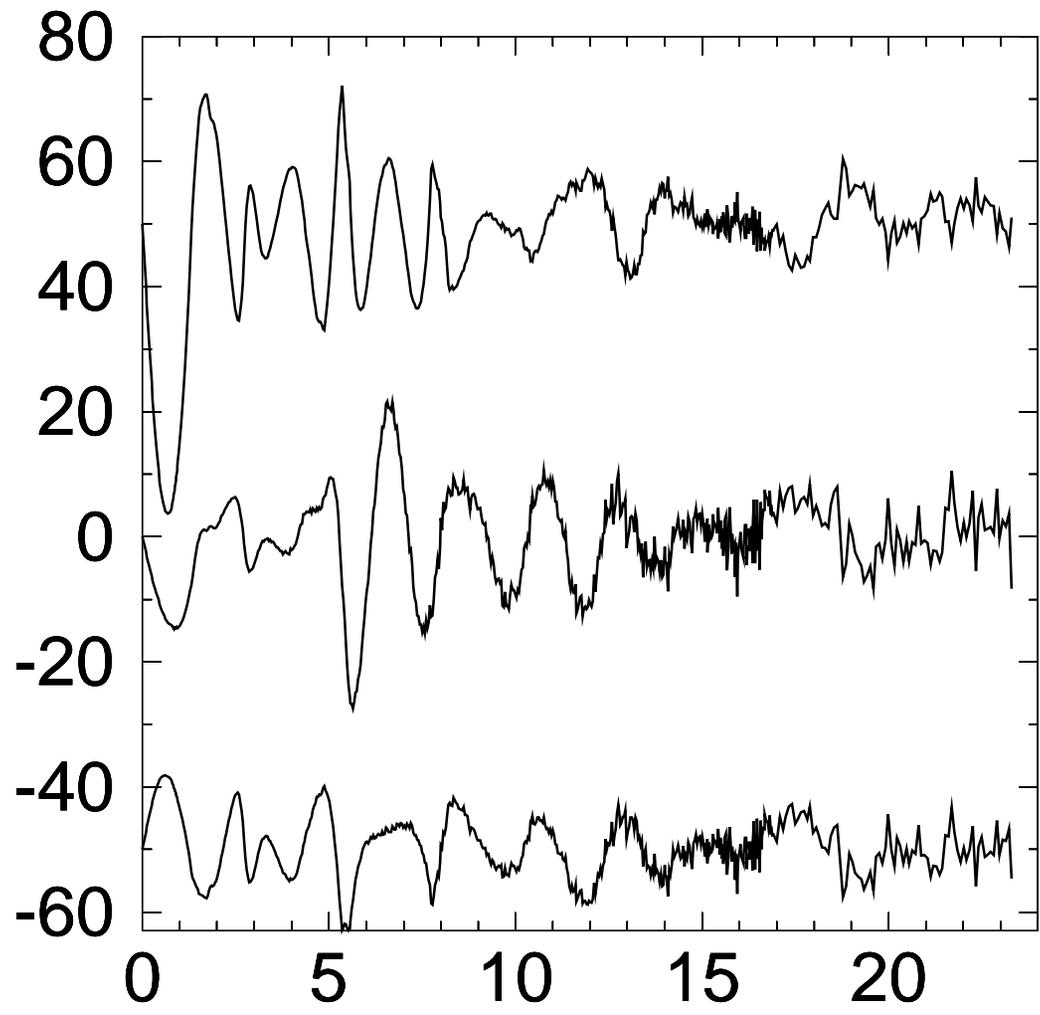
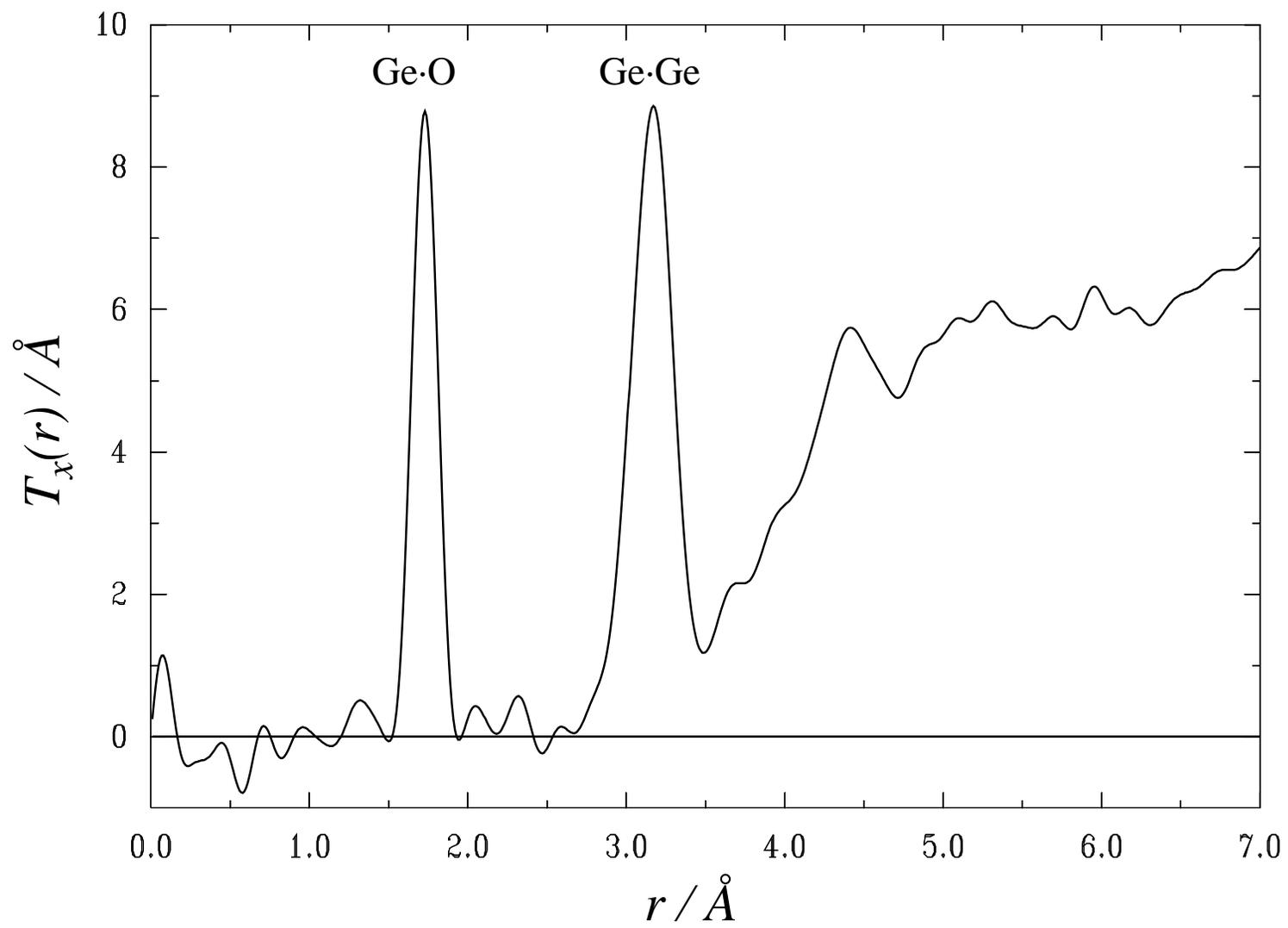
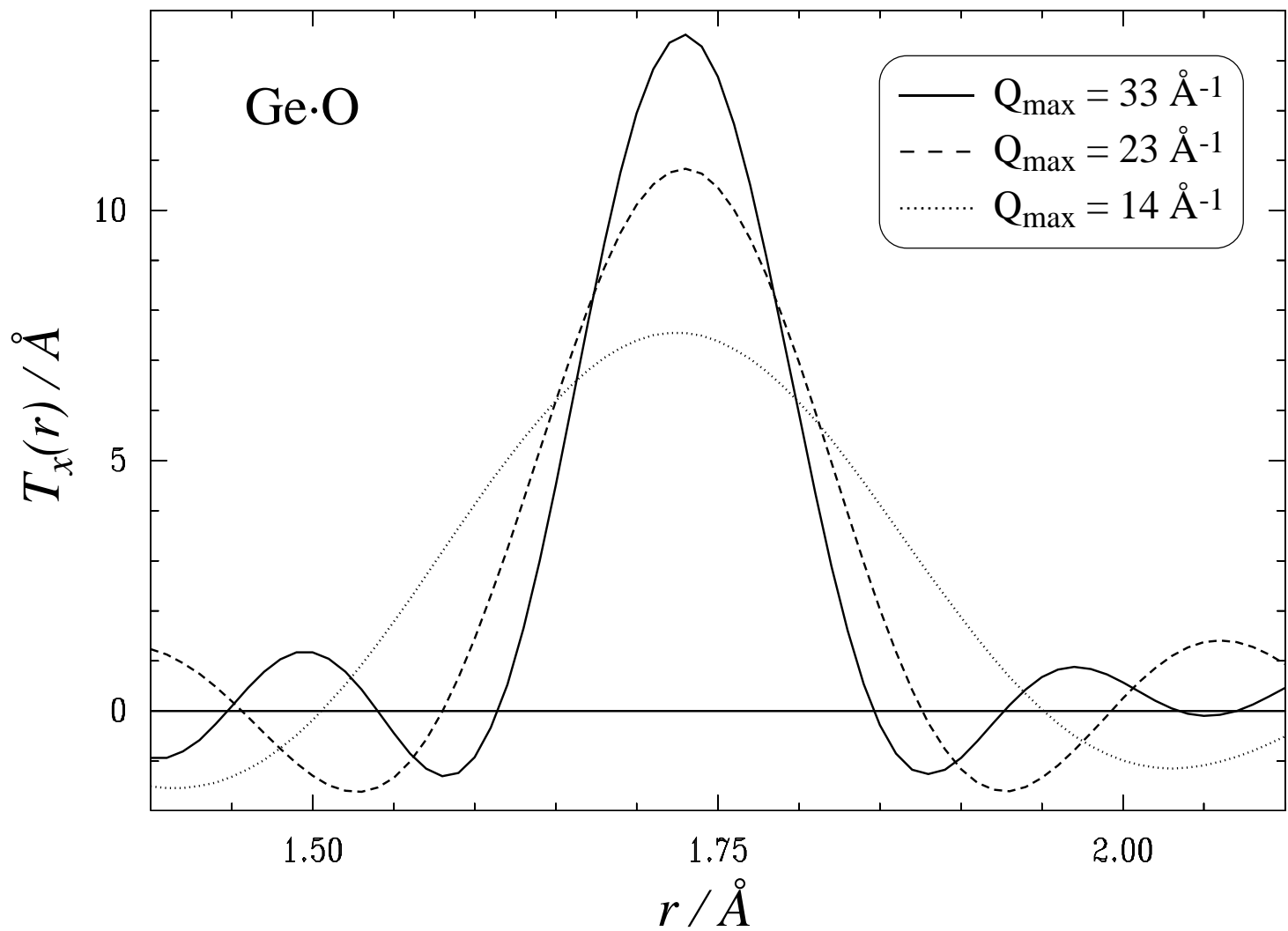
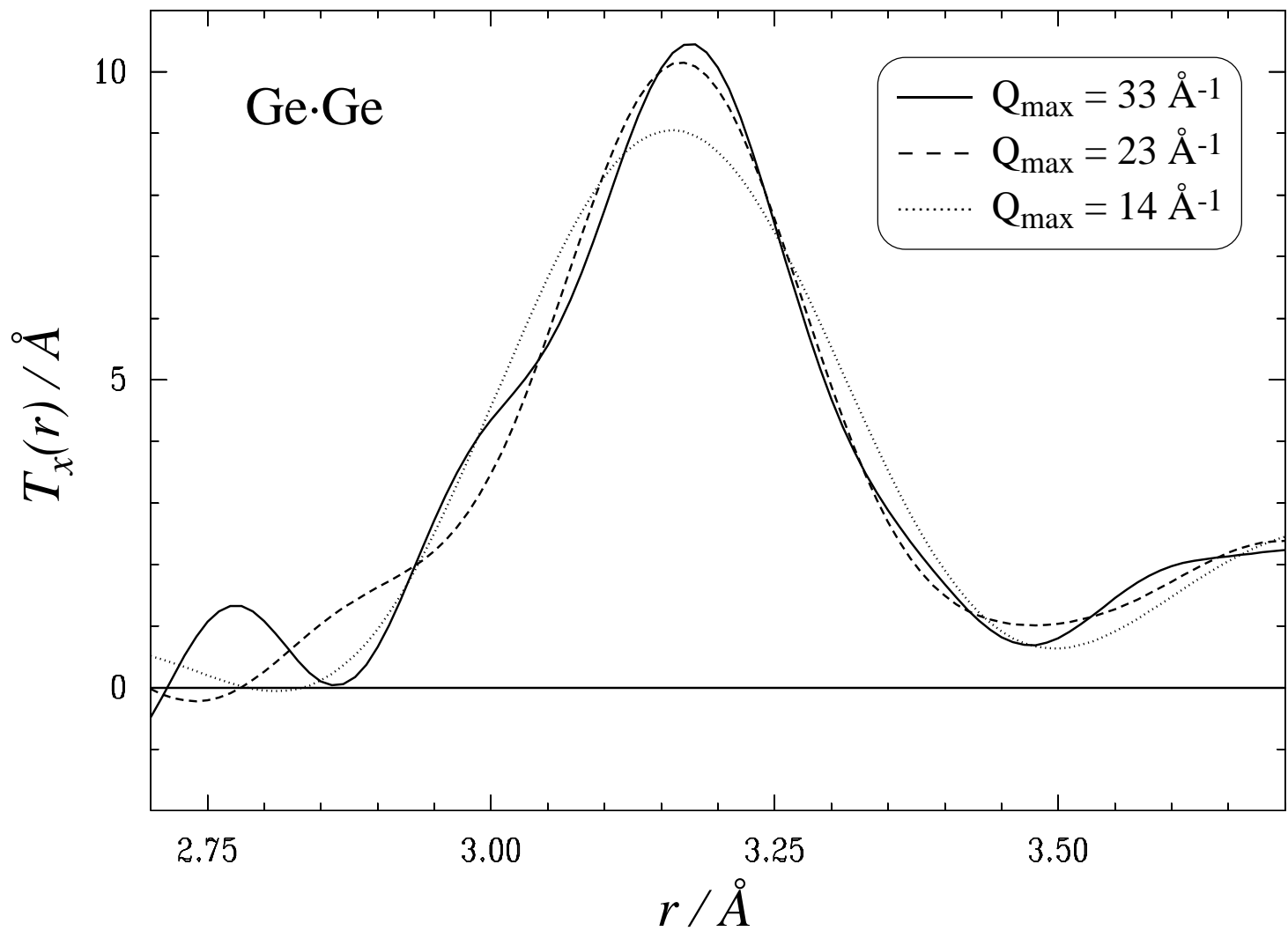
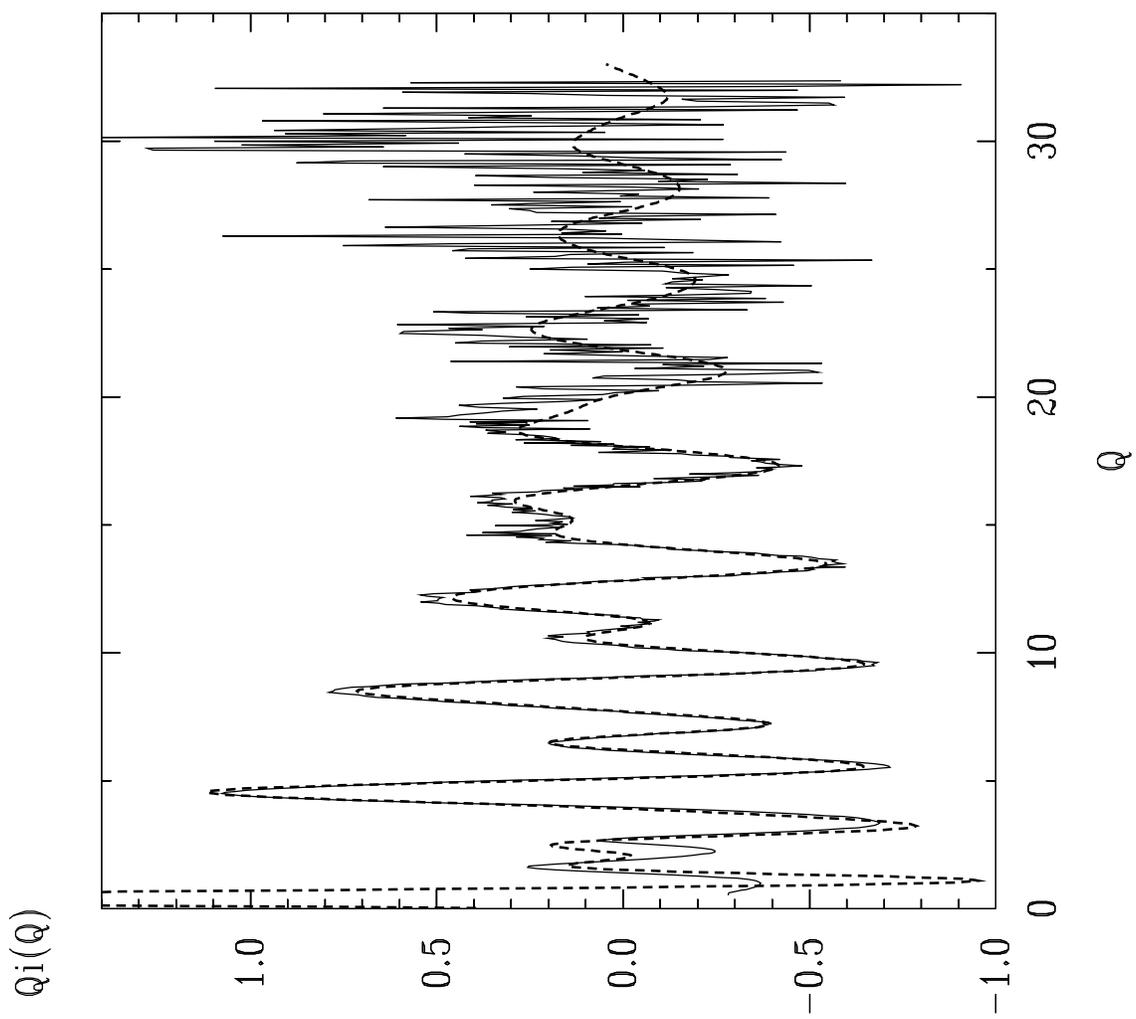
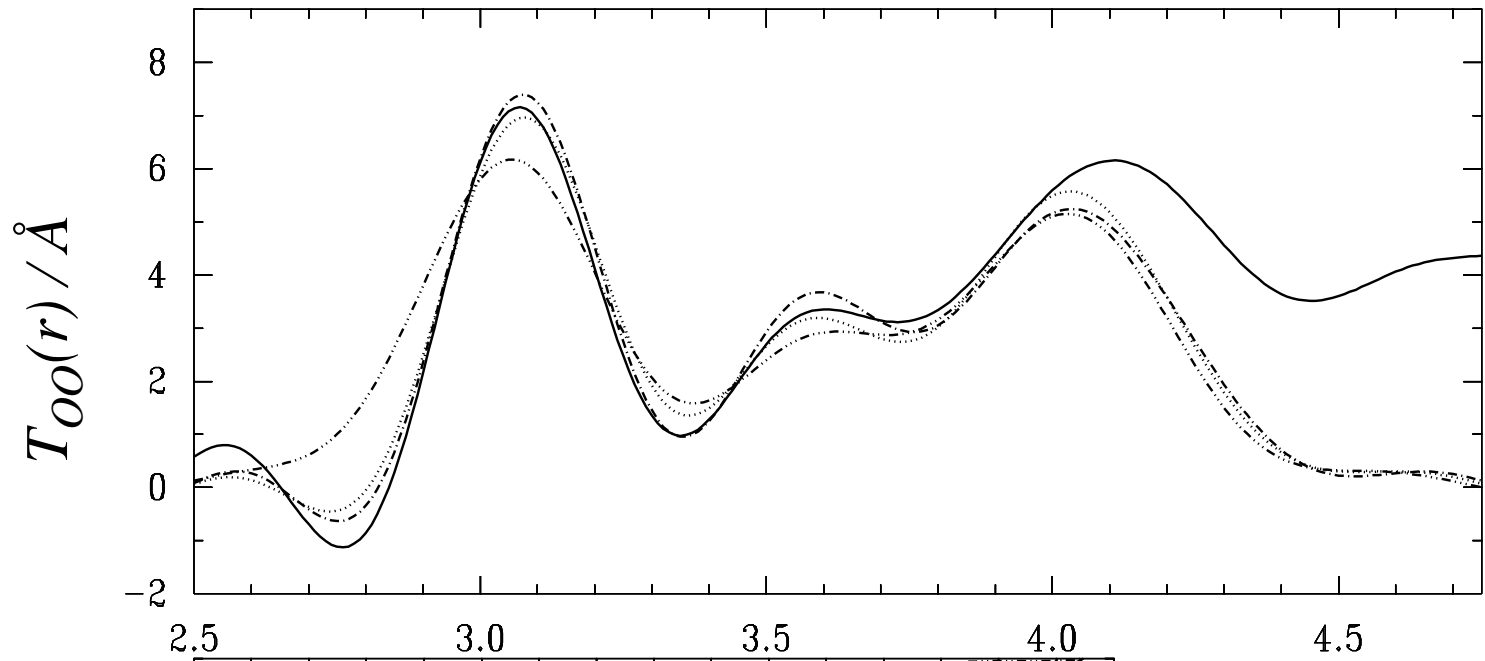
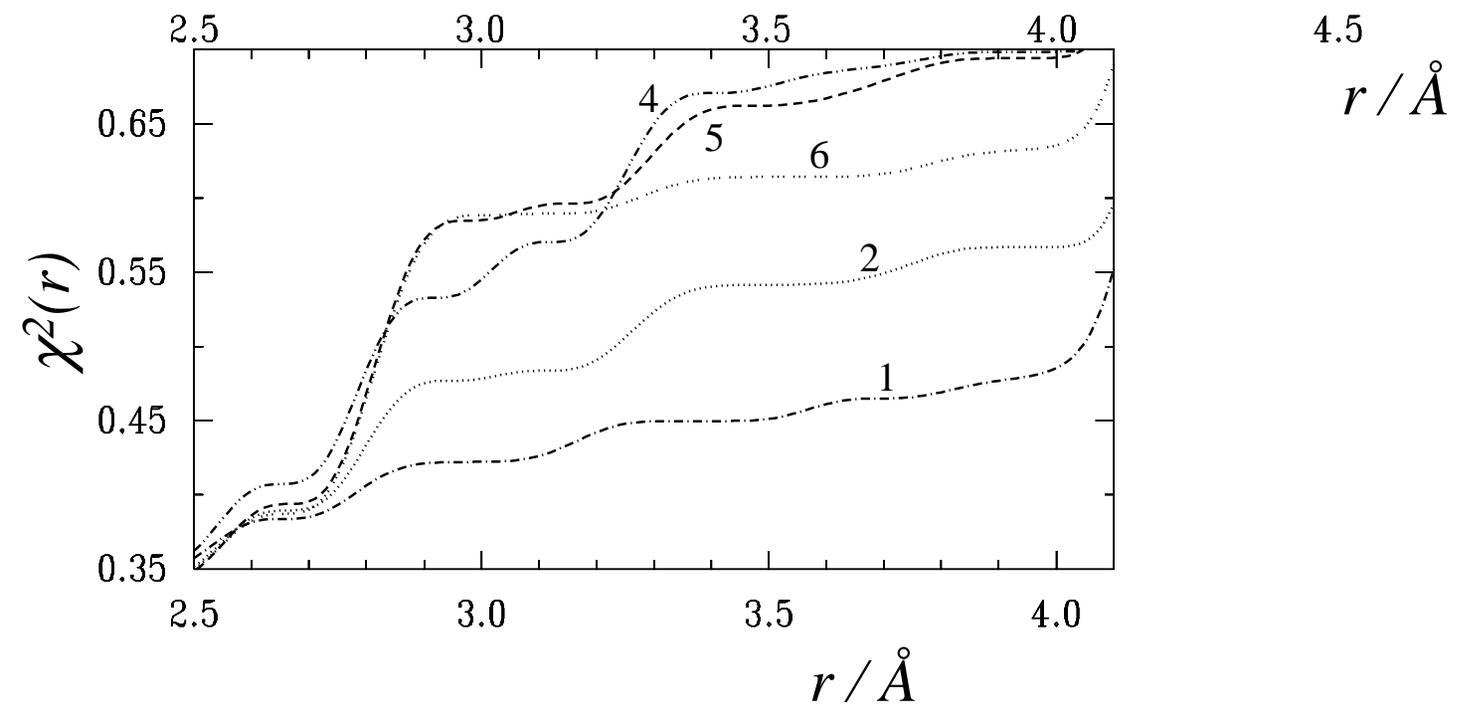
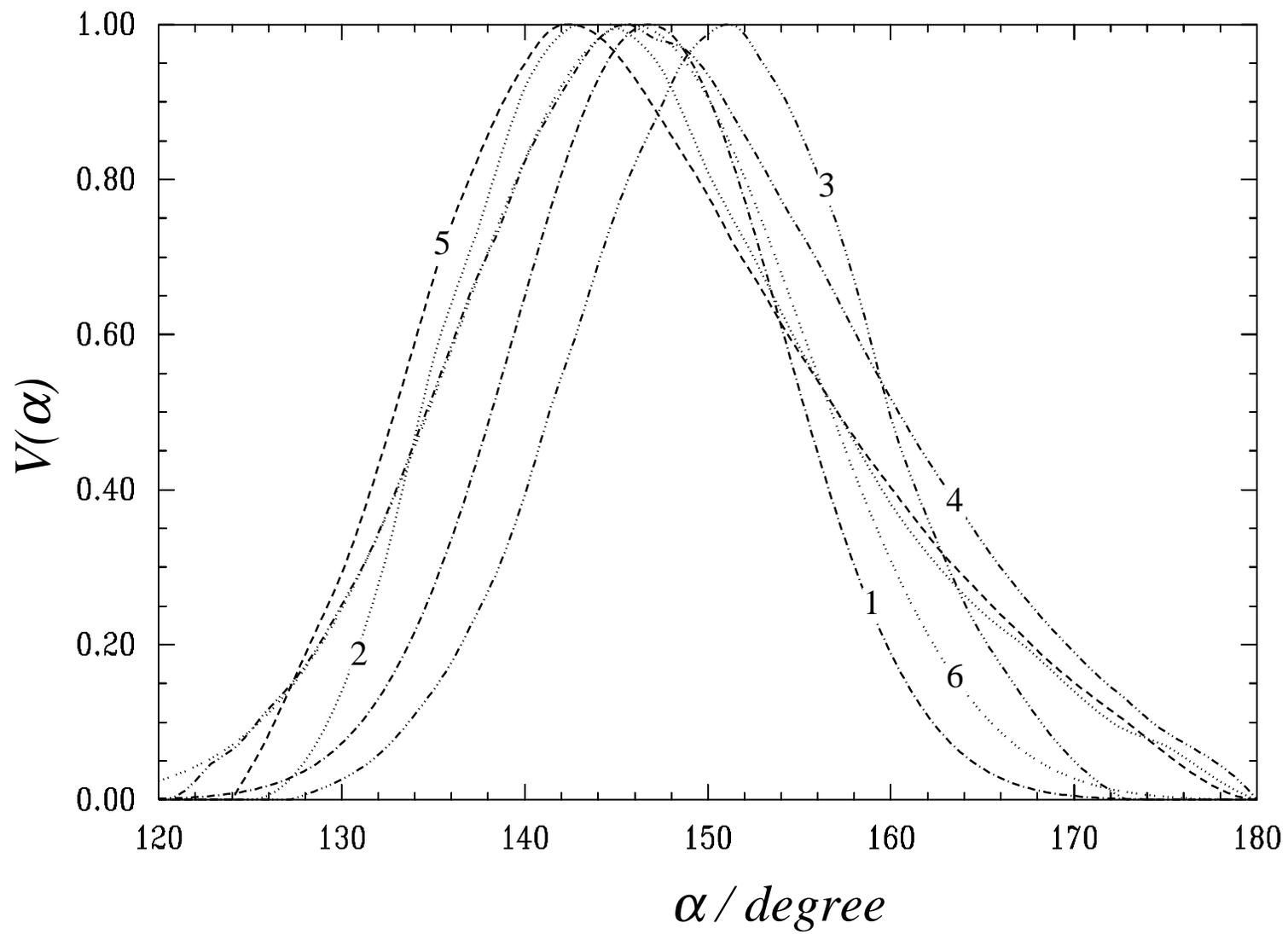
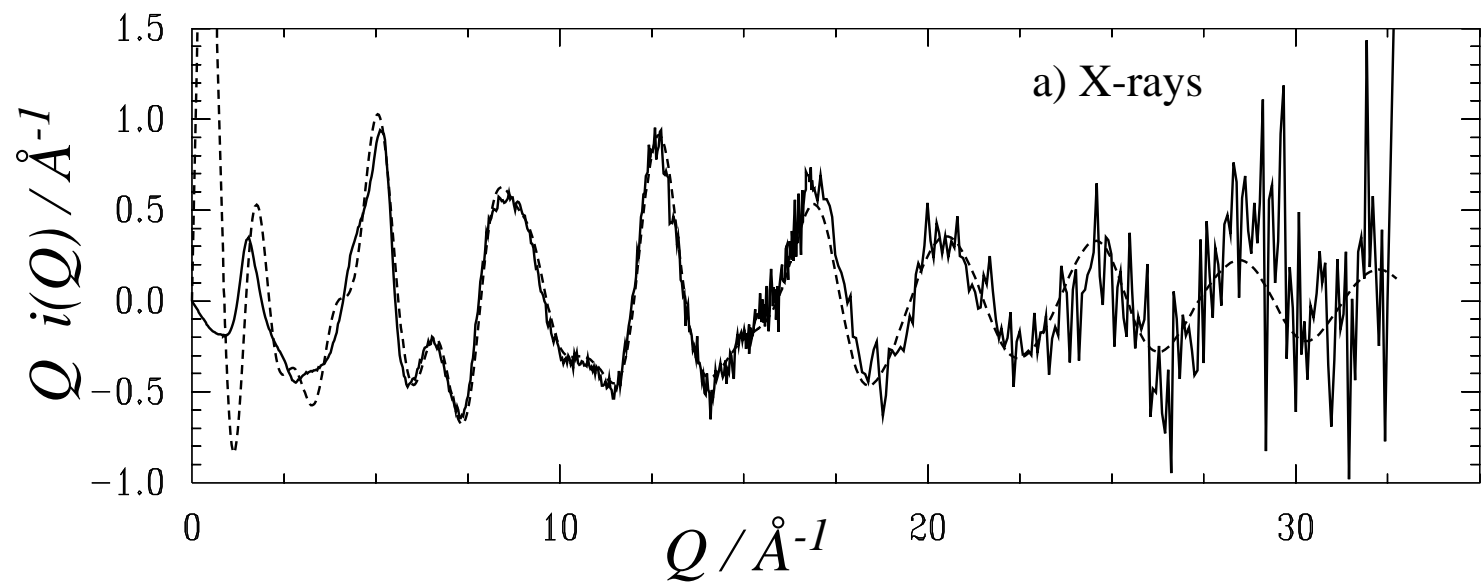
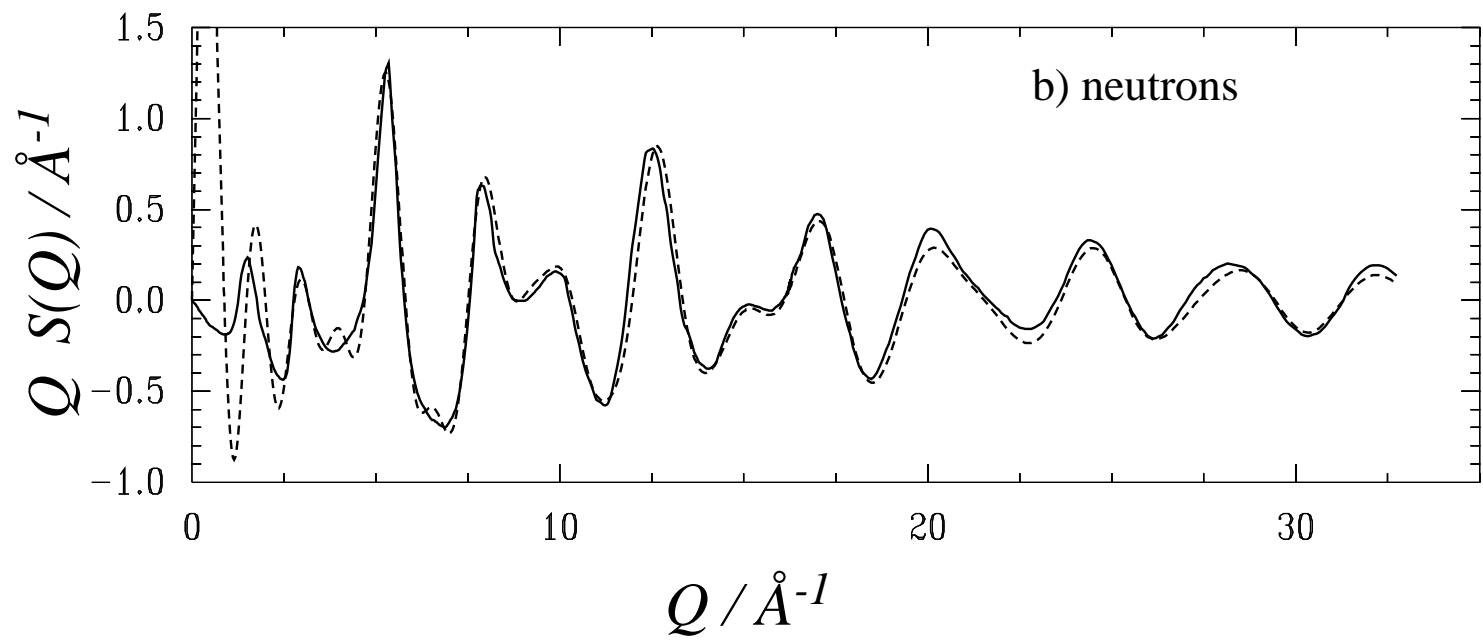
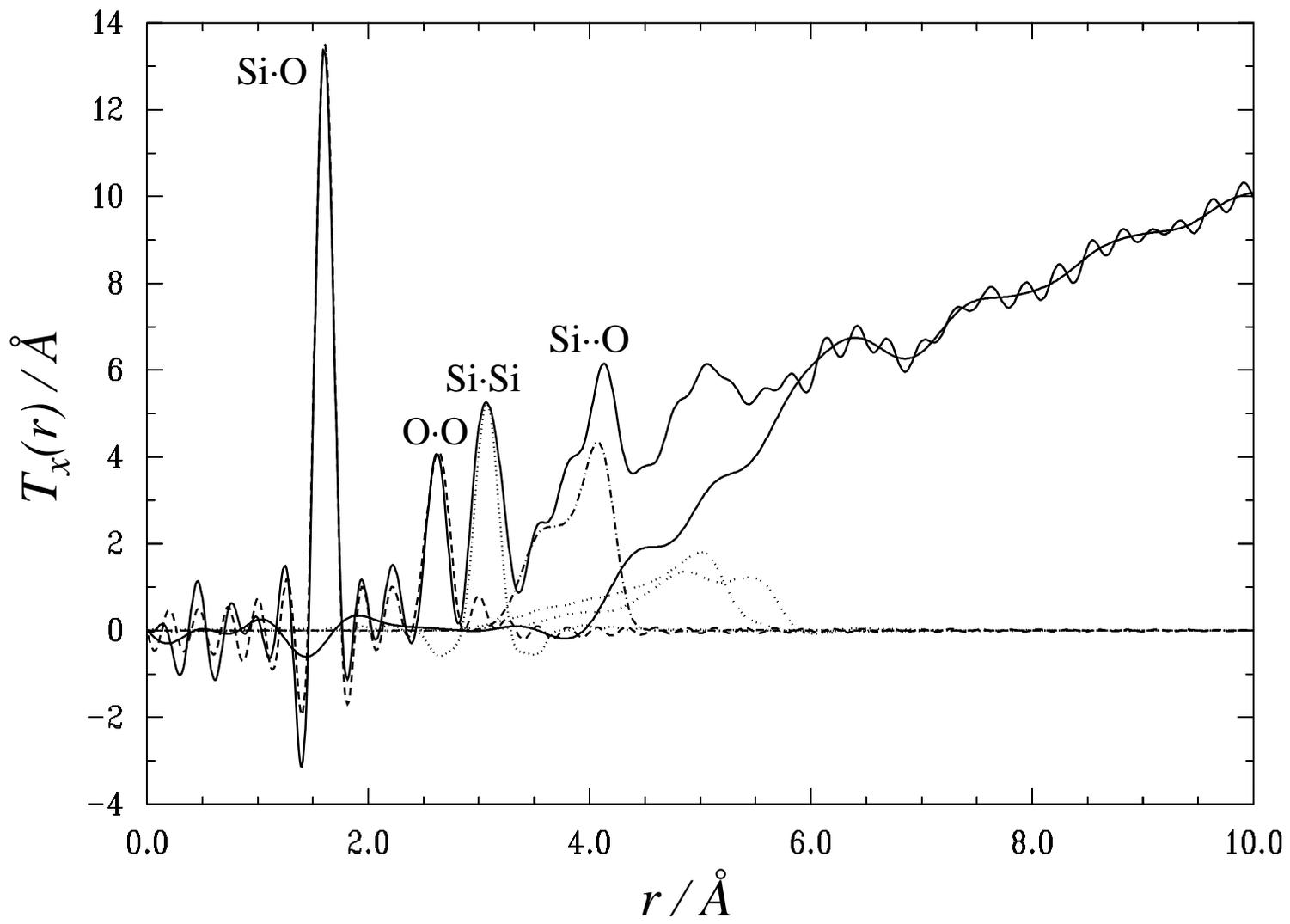
